\documentclass[a4paper,11pt]{article}
\usepackage{pos}
\usepackage{hyperref}
\usepackage{lineno}  % Add this package

\title{Study of Resonance Production using Run 3 pp Collisions with ALICE}
\author*[1,\dagger]{Hirak Kumar Koley}

\affiliation[\dagger]{Nuclear and Particle Physics Research Centre,\\
   Department of Physics, Jadavpur University, Kolkata - 700032, INDIA}

\emailAdd{hirak.koley@cern.ch}

\abstract{
\par{Recent measurements in small collision systems at the LHC show striking similarities between high multiplicity pp, p--Pb collisions and Pb--Pb collisions. In particular, study of hadronic resonances provide valuable information about the final state hadronic interactions. Due to their short lifetime, resonances decay inside the hadronic medium after the chemical freezeout and their decay daughters interact elastically with other hadrons. As a consequence, measured resonance yields get modified. The ALICE experiment is suitable for measuring hadronic resonances thanks to its excellent tracking and particle identification capabilities over a broad momentum range. In this contribution, new measurements of $\mathrm{K}(892)^{*0}$, $\phi(1020)$, and $\Lambda(1520)$ resonance production using high statistics pp collisions at $\sqrt{s} =$ 0.9 and 13.6 TeV collected by the ALICE Collaboration during the Run 3 data taking are presented.}
}

\FullConference{42nd International Conference on High Energy Physics (ICHEP2024)\\
18-24 July 2024\\
Prague, Czech Republic\\}

\begin{document}
%\linenumbers  % Enable line numbers
\footnotetext[1]{On behalf of the ALICE collaboration} % Text of the footnote
\maketitle
%\section*{Introduction}
The study of resonance states plays a pivotal role in understanding the hadronic evolution and final-state interactions in high energy hadronic and nuclear collisions. Resonances have lifetimes in the order of ${\sim}\mathrm{fm}\it{/c}$, comparable to the hadronic phase lifetime, making them highly sensitive to re-scattering and regeneration processes occurring between chemical and kinetic freeze-outs \cite{KOCH1986167}. The ratios of resonance yields to their ground state counterparts exhibit significant variation with multiplicity. Notably, resonances such as $\mathrm{K}^{*0}$ and $\Lambda(1520)$ are strongly suppressed in high-multiplicity events, signifying their sensitivity to the interactions in the hadronic medium. In contrast, the $\phi(1020)$ resonance does not show any suppression, indicating a distinctly different interaction mechanism. This behavior evolve smoothly across various collision systems, suggesting a consistent underlying pattern in resonance production and suppression dynamics, which provides key insights into the hadronic interactions.
\paragraph{}
The extensive upgrade of the ALICE detector during the second long shutdown (LS2) of the LHC (2018–2021) enhance its vertexing, tracking, readout and luminosity capabilities \cite{upgrades}. The ALICE detector consists of several advanced subsystems optimized for tracking and particle identification. The Time Projection Chamber (TPC) serves as the primary tracking device, offering exceptional spatial resolution and particle identification through specific ionization energy loss (dE/dx) measurements. Complementing the TPC, the Inner Tracking System (ITS) provides precise vertex reconstruction and enhances low-momentum particle tracking and identification. In Run 3, the TPC features a new design utilizing Gas Electron Multiplier (GEM) technology, while the upgraded ITS comprises seven cylindrical layers of Monolithic Active Pixel Sensors (MAPS), with the innermost layer located just 23 mm from the interaction point. Additionally, the Time-of-Flight (TOF) detector aids particle identification by measuring particle flight times with high precision.
In Run 3, ALICE can now handle an interaction rate of 50 kHz, resulting in a 100-fold increase in statistics for minimum bias triggers and a 10-fold increase for rare muon triggers, compared to the previous program in Run 2. This upgrade significantly boosts the statistical precision of key measurements, enabling deeper exploration into QGP dynamics and exotic resonance production.
\paragraph{}
The Run 3 of the LHC was officially inaugurated in July 2022 with proton-proton (pp) collisions. Since then, ALICE collected an unprecedented dataset of over 65 $\mathrm{pb}^{-1}$ in pp collisions at $\sqrt{s} =$ 13.6 TeV. Prior to that a test run was done with $\sqrt{s} =$ 0.9 TeV. At 0.9 TeV, the lowest multiplicity region can be accessed, providing an unique opportunity to investigate the canonical suppression of strangeness, offering the first-ever measurements of the $\mathrm{K}^{*0}$ and $\Lambda(1520)$ resonances in this environment. These measurements offer new insights into resonance production and the underlying mechanisms in low-multiplicity collisions. On the other hand, 13.6 TeV data is by far the largest dataset available, facilitating comprehensive differential resonance studies. These include investigations of hadron-resonance correlations and resonance flow, significantly enhancing our understanding of particle interactions in high-energy collisions. 

\begin{figure}[!h] \centering \begin{minipage}{0.49\textwidth} \centering \includegraphics[width=3.67cm,height=3.9cm]{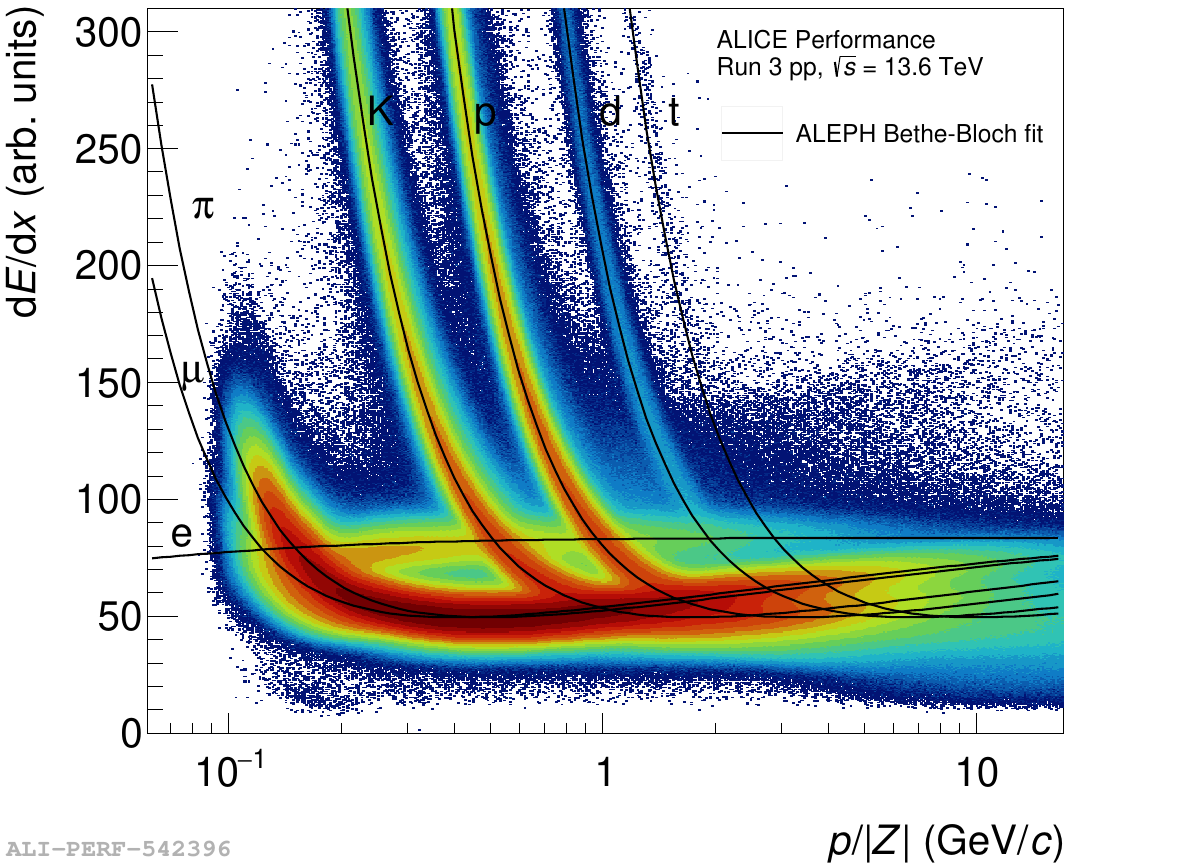} \includegraphics[width=3.65cm,height=4.04cm]{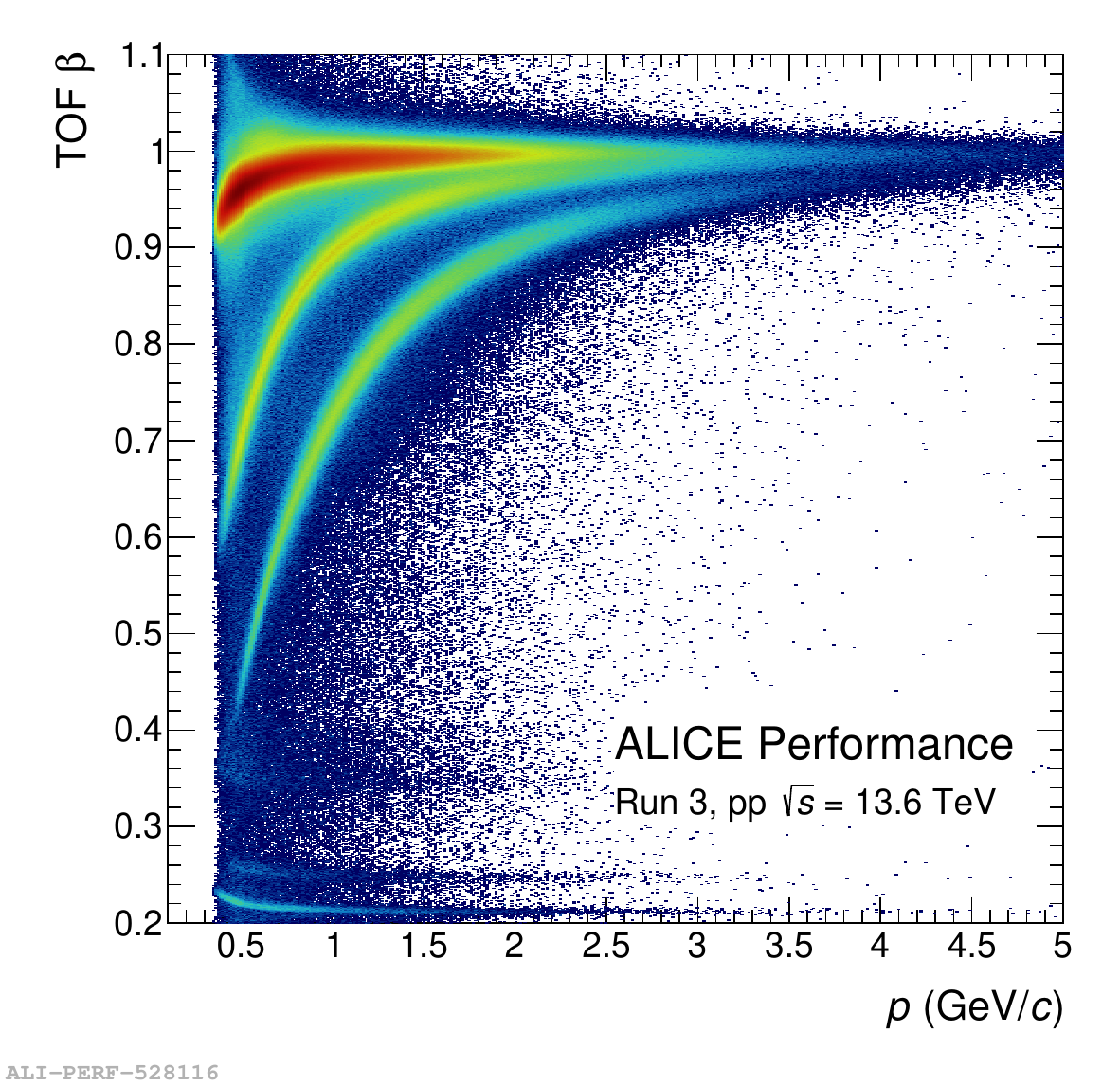} \caption{\label{fig1} TPC dE/dx (left) and TOF $\beta$ (right) distributions for charged particles in pp collisions at $\sqrt{s} =$ 13.6 TeV.} \end{minipage} \hfill \begin{minipage}{0.49\textwidth} \centering \includegraphics[scale=0.3]{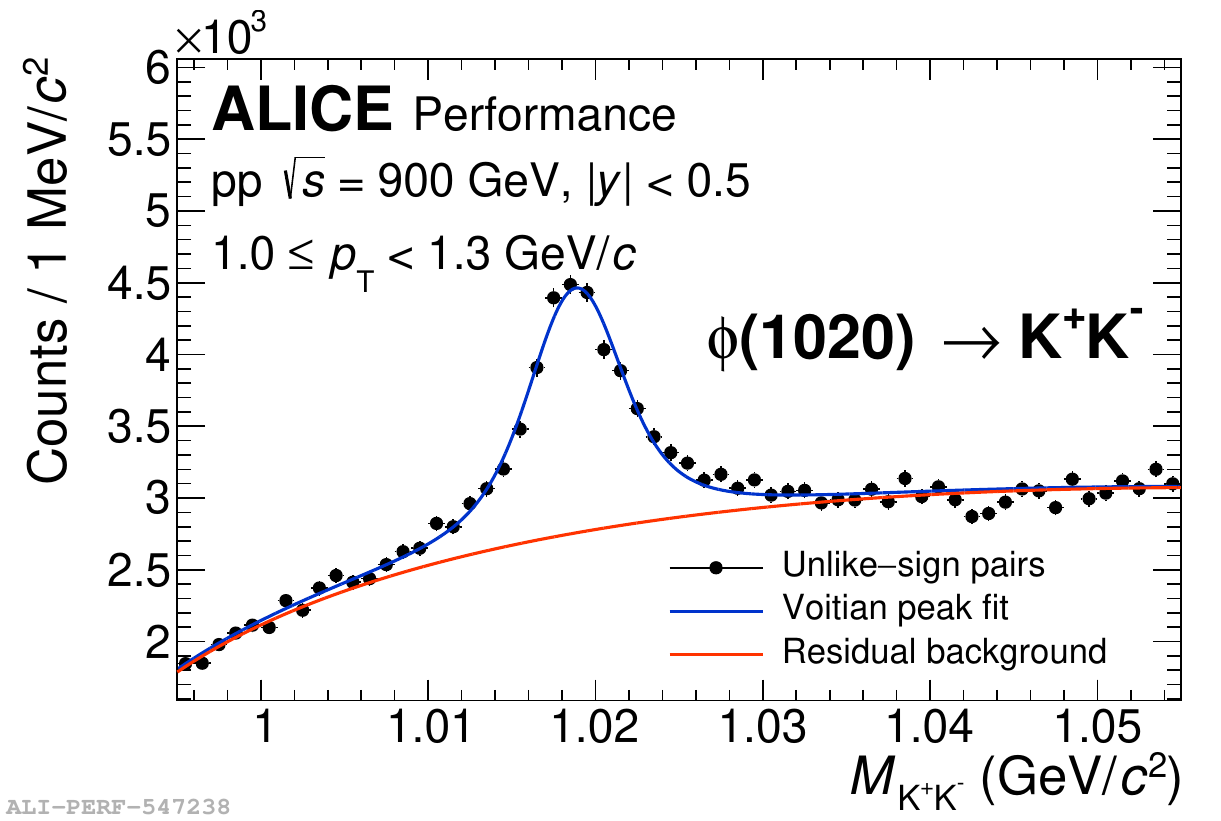} \caption{\label{fig2} Invariant mass distribution for $\phi(1020)$ in pp collisions at $\sqrt{s} =$ 0.9 TeV.} \end{minipage} \end{figure}

%\section*{Performance of Resonance Reconstruction in Run 3}
\paragraph{}
ALICE has successfully measured $\mathrm{K}^{*0}$, $\phi(1020)$, and $\Lambda(1520)$ resonances by reconstructing their daughter tracks within the rapidity range $|y| < 0.5$. This analysis focuses on their production in proton-proton (pp) collisions at two center-of-mass energies: $\sqrt{s} =$ 0.9 TeV and 13.6 TeV. Events were selected using a minimum bias trigger, with the additional requirement that the reconstructed primary vertex lies within $\pm$10 cm of the nominal interaction point along the beam axis. Particle identification in this study primarily utilizes the TPC and the TOF detector. Figure \ref{fig1} illustrates the TPC dE/dx and TOF $\beta$ distributions for charged particles in pp collisions at $\sqrt{s} =$ 13.6 TeV.

\begin{figure}[!h] \centering \begin{minipage}{0.49\textwidth} \centering \includegraphics[width=3.66cm,height=5.55cm]{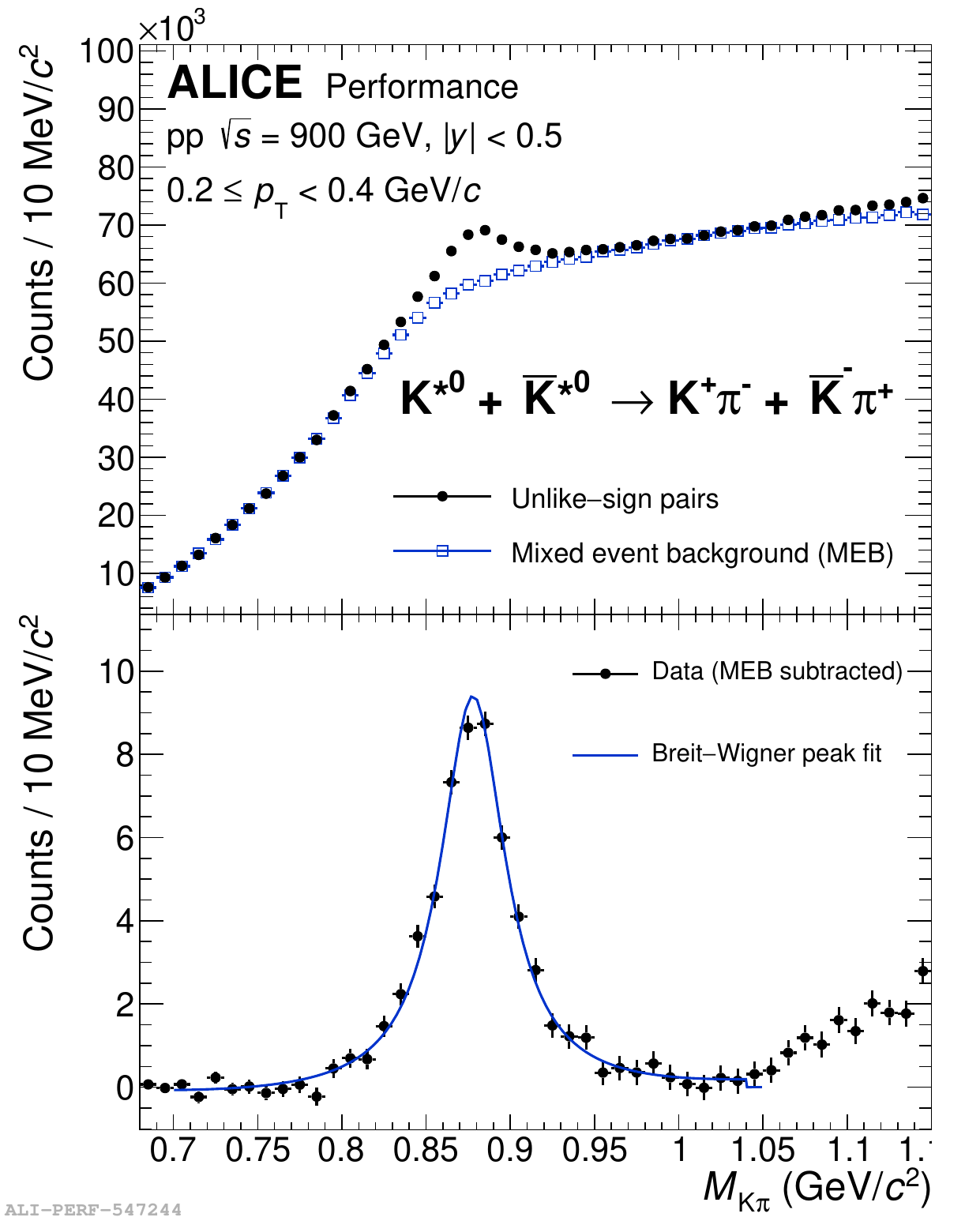} \includegraphics[width=3.66cm,height=5.55cm]{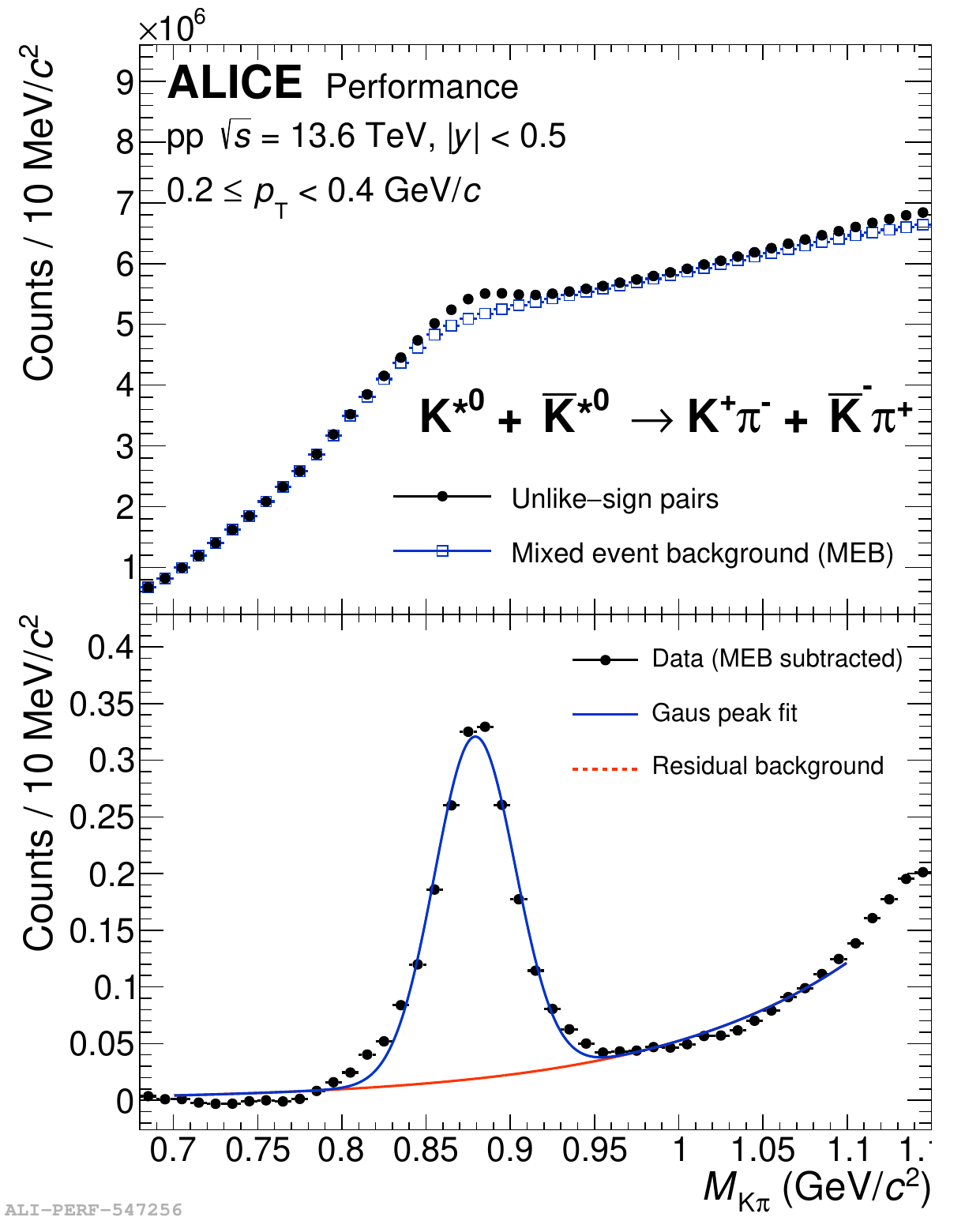} \caption{\label{fig3} Invariant mass distributions for $\mathrm{K}^{*0}$ at $\sqrt{s} =$ 0.9 and 13.6 TeV.} \end{minipage} \hfill \begin{minipage}{0.49\textwidth} \centering \includegraphics[width=3.66cm,height=5.55cm]{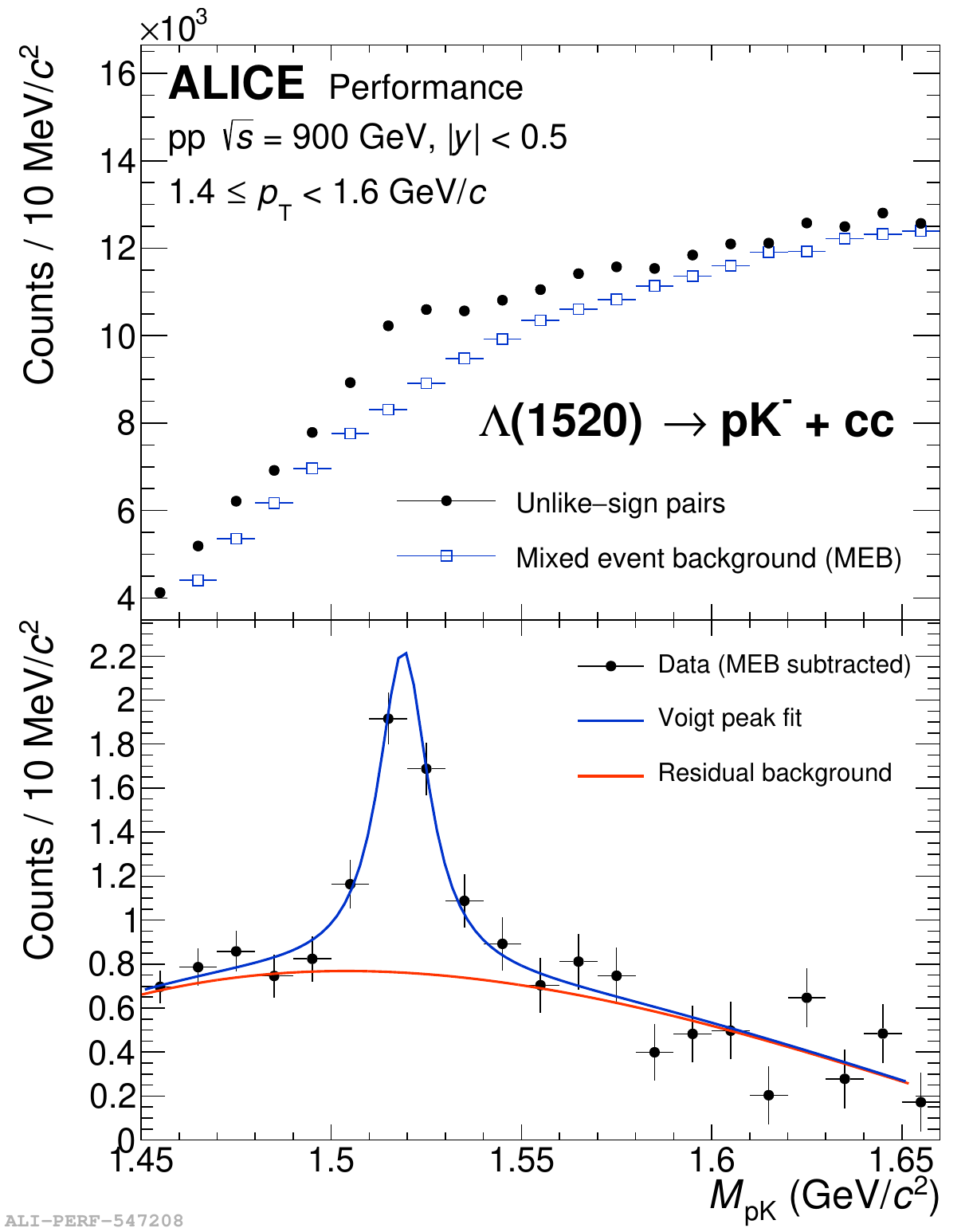} \includegraphics[width=3.66cm,height=5.55cm]{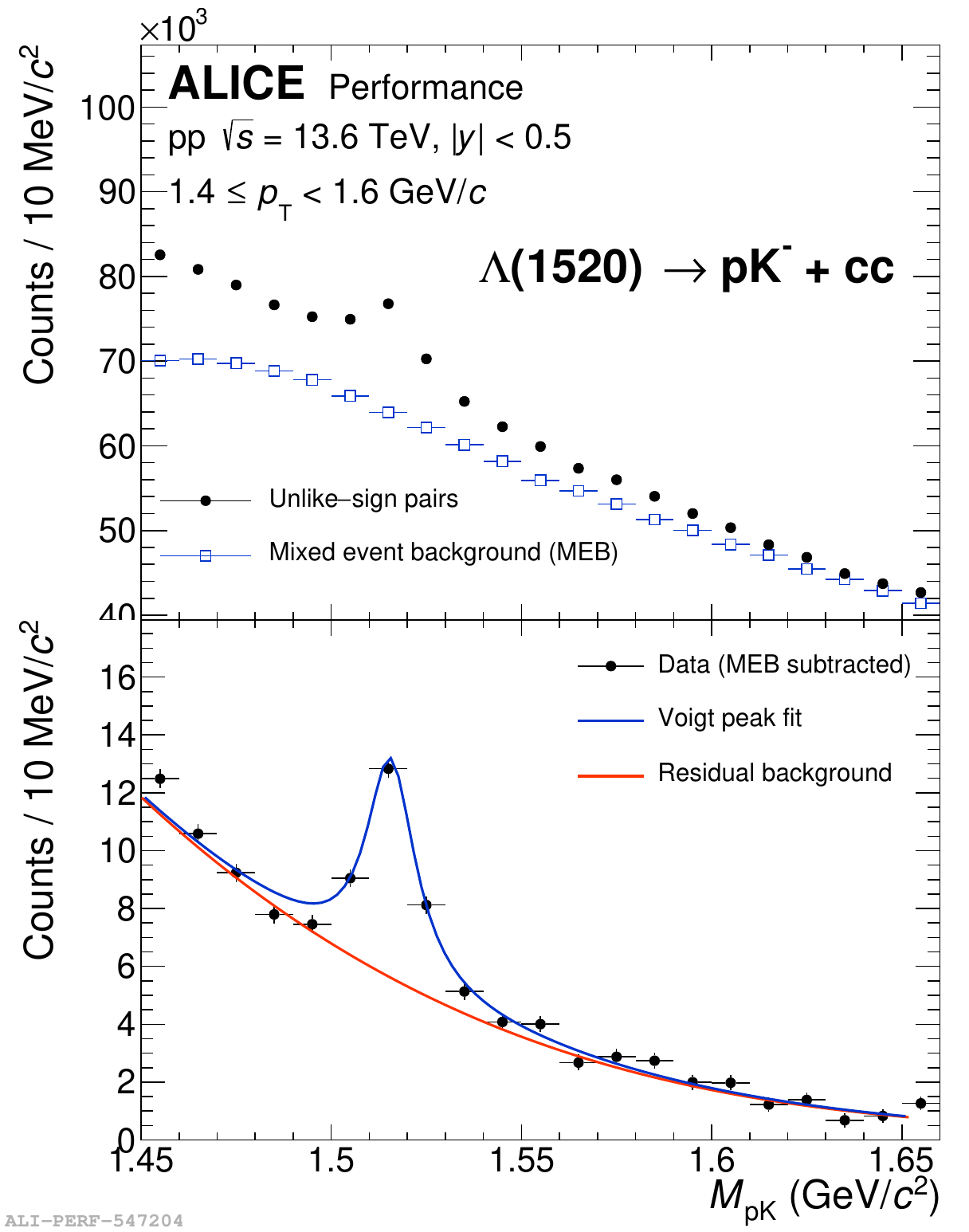} \caption{\label{fig4} Invariant mass distributions for $\Lambda(1520)$ at $\sqrt{s} =$ 0.9 and 13.6 TeV.} \end{minipage} \end{figure}

%\paragraph{}
\vspace{-0.5em}
In Figure \ref{fig2}, the invariant mass distribution for the $\phi(1020)$ resonance is displayed, reconstructed through the $\mathrm{K}^{\pm}\mathrm{K}^{\mp}$ decay channel in pp collisions at $\sqrt{s} =$ 0.9 TeV. Similarly, Figures \ref{fig3} and \ref{fig4} show invariant mass distributions for $\mathrm{K}^{*0}$ and $\Lambda(1520)$, respectively, using the $\mathrm{K}^{\pm}\pi^{\mp}$ and $\mathrm{pK}^{-}(\Bar{\mathrm{p}}\mathrm{K}^{+})$ decay channels, reconstructed from pp collisions at $\sqrt{s} =$ 0.9 TeV and 13.6 TeV.

%\paragraph{}
\vspace{-0.5em}
These invariant mass distributions provide clear signals for the resonances, further detailed analyses are currently underway, including other resonances. We anticipate that ongoing studies will contribute substantially to our understanding of the role of resonances in probing QGP properties and final-state interactions in various collision systems.

\paragraph{} The author, Hirak Kumar Koley, gratefully acknowledges financial support from the Department of Science and Technology, Government of India, under the “Mega Facilities in Basic Science Research” scheme.
%\end{acknowledgments}

%\ \\
%\noindent

\end{document}